\documentclass[prl,twocolumn]{revtex4}
\usepackage{graphicx}
\usepackage{amsmath,amssymb}
\DeclareMathOperator{\Tr}{Tr}
\begin{document}
\title{Fast and accurate quantum molecular dynamics \\of dense plasmas across temperature regimes}
\author{Travis Sjostrom and J\'er\^ome Daligault}
\affiliation{Theoretical Division, Los Alamos National Laboratory,
Los Alamos, New Mexico 87545}
\date{\today}
\begin{abstract}
We have developed and implemented a new quantum molecular dynamics approximation that allows fast and accurate simulations of dense plasmas from cold to hot conditions. The method is based on a carefully designed orbital-free implementation of density functional theory (DFT). The results for hydrogen and aluminum are in very good agreement with Kohn-Sham (orbital-based) DFT and path integral Monte Carlo (PIMC) for microscopic features such as the electron density as well as equation of state. The present approach does not scale with temperature  and hence extends to higher temperatures than is accessible in Kohn-Sham method and lower temperatures than is accessible by PIMC, while being significantly less computationally expensive than either of those two methods.
\end{abstract}
\maketitle

A significant challenge of high energy density physics is the determination of the fundamental properties of plasmas (e.g. equation of state, transport properties) over a wide range of temperatures and densities \cite{Drake,HEDLP}. Systems of particular focus include  warm dense matter \cite{IPAM}, inertial confinement fusion, notably the compression pathway to ignition, and astrophysical plasmas. Two methods have emerged as standards for such calculations which have yielded quality results. Those are Kohn-Sham density functional theory based molecular dynamics \cite{Mazevetetal,Holstetal,Wangetal} and path integral Monte Carlo \cite{Huetal,DriverMilitzer}. Due to the nature of the method, PIMC becomes prohibitive as the temperature is decreased and Kohn-Sham DFT becomes prohibitive with increasing temperature as the number of required orbitals increases with temperature and in general the method scales as the cube of the number of orbitals. It is possible to find the region of overlap for these calculations, but such a region is generally difficult for both methods \cite{DriverMilitzer,Militzer09}. In this letter we develop and implement an orbital-free DFT formulation which provides accuracy at the level of Kohn-Sham DFT and PIMC while spanning from low to high temperatures, without any scaling with temperature and at significantly lower computational cost than the other two methods.

In DFT the fundamental quantity is the free energy, which is minimized to find the electron density. For a given ionic configuration the free energy is a functional of the electron density, $n$, and is given by  \cite{ParrYang}
\begin{equation}
  F[n]=F_s[n] + F_H[n] + F_{xc}[n] + F_{ei}[n] 
  \label{eq:fe1}
\end{equation}
where $F_s$ is the non-interacting free energy comprised of both kinetic and entropic parts, $F_H$ is the Hartree energy or direct Coulomb interaction between the electrons, $F_{\rm ei}$ is the electron-ion Coulomb interaction, and $F_{xc}$ is defined as the remainder of the total free energy, which includes the quantum mechanical exchange and correlation as well as the excess kinetic and entropic terms. Of the contributions neither $F_s$ or $F_{xc}$ have explicitly calculable forms. Given the same orbital-free $F_{xc}$ approximation, the only difference in approach of orbital-free DFT from Kohn-Sham DFT is that the non-interacting free energy, $F_s$, is approximated by a density functional instead of being exactly obtained through the calculation of single particle orbitals \cite{KohnSham}. Thus returning to a pure DFT which, as given by the Hohengberg-Kohn-Mermin theorems \cite{HohenbergKohn,Mermin}, is an exact theory. 

Significant efforts have been made at zero temperature in developing advanced orbital-free functionals with high quality results \cite{WangTeter,SmargiassiMadden,Perrot,WGC1,WGC2,HuangCarter,Xiaetal,Gonzalezetal}. Though without analogous efforts, in recent years the orbital-free approach at finite temperature has gained attention, with most results being for hot dense systems where the venerable Thomas-Fermi approximation is employed for $F_s$ \cite{Lambert06,Burakovskyetal,Clerouinetal}. The work of Perrot offered a density gradient correction to Thomas-Fermi that improves results moderately \cite{Perrot79,Daneletal}. Other more recent semi-local functionals \cite{ftgga,ftgga2} have also been considered. None of these functionals, though,  have reached the accuracy of Kohn-Sham across temperature regimes.

In this work we develop and implement an advanced  density functional for $F_s$, valid at zero temperature as well as finite temperature, which provides highly accurate agreement with the Kohn-Sham results. Our $F_s$ does not scale with the temperature as the Kohn-Sham method does and is significantly less computationally expensive at low temperatures, since the dependence is on the density only.

We now give a summary of our functional. Further details of individual terms, and all other quantities necessary for quantum molecular dynamics implementation are given in the Supplemental Material \cite{SuppMatt}. The proposed functional is of the following form for the non-interacting free energy
\begin{align}
  F_s[n] = F_{TF}[n] + {}_{\beta}F_{vW}[n] + F_{a,b}[n].
\end{align}
Here the first term on the RHS is the familiar Thomas-Fermi term
\begin{equation}
  F_{TF}[n] = \int f_{TF}(n(\mathbf{r})) \;d{\mathbf r} \;,
  \label{eq:tf}
\end{equation}
where $f_{TF}$ is just the non-interacting electron gas energy per volume at density $n$. The second term on the RHS is the here proposed extension of the semi-local von Weisz\"acker term
\begin{align}
 {}_{\beta}F_{vW}[n] = \frac{\hbar^2}{2m_e} \iint& [(\nabla n^{1/2}(\mathbf{r}))\cdot(\nabla n^{1/2}(\mathbf{r}^{\prime}))] \times \nonumber \\
& [\delta({\mathbf r}-{\mathbf r}^{\prime})+\beta(|{\mathbf r}-{\mathbf r}^{\prime}|)]\; d{\mathbf r}^{\prime}d{\mathbf r}\;.
  \label{eq:betavw}
\end{align}
In the limit $\beta(|{\mathbf r}-{\mathbf r}^{\prime}|)=0$ this reduces to the standard von Weisz\"acker term
\begin{equation}
  F_{vW}[n] = \frac{\hbar^2}{m_e} \int \frac{|\nabla n(\mathbf{r})|^2}{8 n(\mathbf{r})} \; d{\mathbf r}\;.
  \label{eq:vw}
\end{equation}
The final term is the previously introduced nonlocal free energy term
\begin{equation}
  F_{a,b}[n] = \iint n^{a}(\mathbf{r}) w(|\mathbf{r}-\mathbf{r^{\prime}}|) n^{b}(\mathbf{r}^{\prime}) \;d\mathbf{r}^{\prime}\;d\mathbf{r} \;.
\label{eq:abnl}
\end{equation}
with $a$ and $b$ free parameters, and chosen to be $a=b=5/6$. 

This leaves still undetermined the kernels $\beta$ and $w$. To proceed the functional is constrained to reproduce the exact density-density response function (Lindhard), $\tilde{\chi}_0$, of the non-interacting uniform electron gas as follows,
\begin{align}
    \tilde{\chi}_{0}^{-1}(k;n_0,T)&=-\hat{\mathrm F}\left( \frac{\delta^2 F_{S}[n,T]}{\delta n(\mathbf{r}) \delta n(\mathbf{r}^{\prime})} \bigg|_{n_0} \right) \;.
    \label{eq:linresp}
\end{align}
Here $\hat{\mathrm F}$ denotes the Fourier transform of the second functional derivative of $F_s$ evaluated at the average density $n_0$. This results in the following relation for the $w$ and $\beta$ kernels in reciprocal space
\begin{align}
  \tilde{w}(k) &= \frac{- \tilde{\chi}_0^{-1}(k) + \tilde{\chi}_{TF}^{-1} + [1+\tilde{\beta}(k)]\tilde{\chi}_{vW}^{-1}(k)}{2abn_0^{(a+b-2)}} \nonumber \\
      &\equiv f(k) \frac{- \tilde{\chi}_0^{-1}(k) + \tilde{\chi}_{TF}^{-1} + \tilde{\chi}_{vW}^{-1}(k)}{2abn_0^{(a+b-2)}}\;,
  \label{eq:w}
\end{align}
where $\tilde{\chi}_{TF}^{-1}$ and $\tilde{\chi}_{vW}^{-1}(k)$ are the contributions to Eq. (\ref{eq:linresp}) from Eq. (\ref{eq:tf}) and Eq. (\ref{eq:vw}) respectively (see Supplemental Material \cite{SuppMatt}). For convenience we have written $\tilde{w}$ in terms of $f(k)$ in the second line.
We may now choose $f(k)$ with the only constraint that $f(k)$ remains finite. 
Satisfaction of Eq. (\ref{eq:w}), then determines $\tilde{w}$  and $\tilde{\beta}$, and guarantees the functional produces the exact response and free energy in the uniform electron gas limit.

\begin{figure}[]
  \includegraphics[width=0.9\columnwidth]{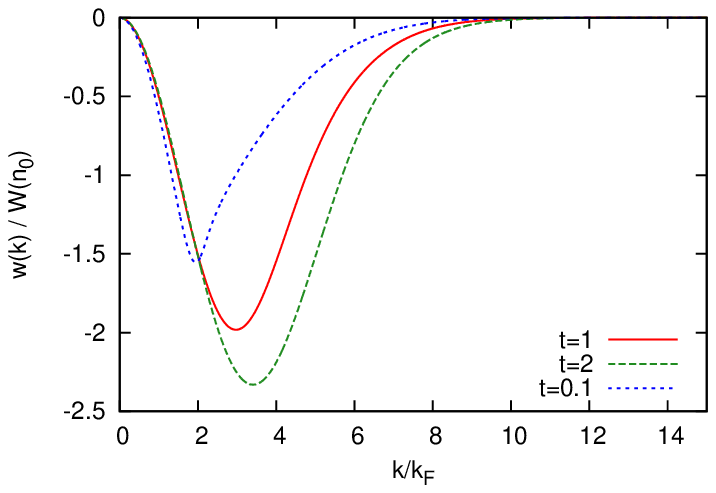}
  \includegraphics[width=0.9\columnwidth]{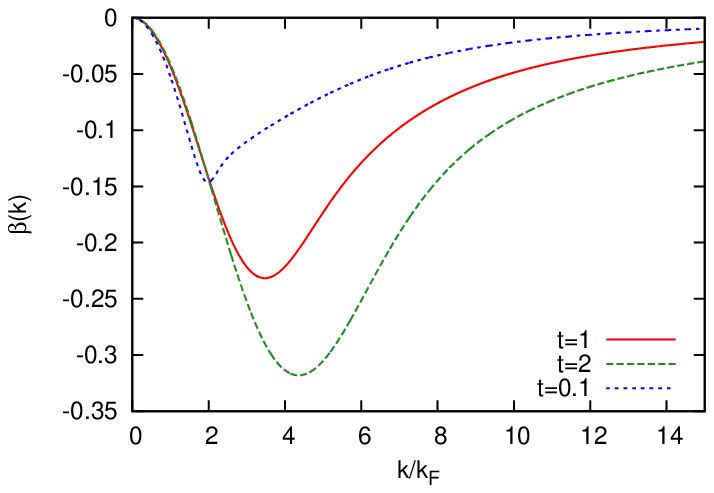}
  \caption{The two kernels of the present functional plotted at several temperatures $t=k_BT/E_F$, with $f(k)=e^{-k^2/4^2 k_F^2}$ which enforces $\tilde{w}(k)$ goes to zero for large $k$. Here $W(n_0)={n_0^{(5/3-a-b)}} \hbar^2 / 6 a b m_e $.}
  \label{fig:nlexp}
\end{figure}

At zero temperature \cite{WangTeter,SmargiassiMadden,Perrot,WGC1,WGC2} and more recently at finite temperature \cite{SjostromDaligault} the case $f\equiv1$ (i.e. $\beta \equiv 0$), has been investigated. Though this case meets the requirement of correcting the response, it produces a kernel, $\tilde{w}$, which goes to constant negative value in the large $k$ limit  ($k > 10 k_F$) and thus results in the functional being unbounded and producing unphysical densities with infinitely negative energy \cite{BlancCances}. We have added the nonlocality $\beta$ in Eq. (\ref{eq:betavw}) to alleviate this issue, while still enforcing the exact response.
In order to force $\tilde{w}(k)$ to zero for large $k$ ($k > 10 k_F$), removing the aforementioned difficulty of the $f\equiv 1$ case, we consider the interpolating $f(k)=e^{-k^2/\alpha^2 k_F^2}$ with $\alpha=4$. 
The resulting kernels are plotted in Fig. \ref{fig:nlexp}. 

We have applied the new functional to hydrogen and aluminum over a wide range of density and temperatures. In these calculations we use a local pseudopotential for all orbital-free calculations as well as for some Kohn-Sham calculations. Using the same pseudopotential provides an apple to apple comparison of our $F_s$ functional to the exact Kohn-Sham method for $F_s$, since we also use the same $F_{xc}$ approximation in all cases. In addition we perform Kohn-Sham calculations with a more standard nonlocal pseudopotential for comparison. 

The details of the calculations are as follows. The local pseudopotentials for hydrogen and aluminum are given in Refs. \onlinecite{ftgga} and \onlinecite{HuangCarterPP} respectively. In  the orbital-free calculations the numeric grid sizes were $64^3$ or $96^3$ depending on system size and density. For the Kohn-Sham calculations we used the Quantum-Espresso code \cite{qespresso} and planewave cutoff energies of 2040 and 680 eV for hydrogen and aluminum respectively, and all calculations were done at the Gamma-point only. All calculations use the local density approximation \cite{pz81} for $F_{xc}$.

\begin{figure}[]
  \includegraphics[width=0.9\columnwidth]{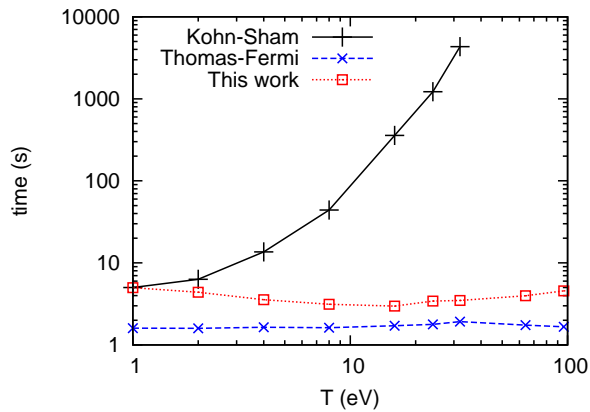}
  \caption{Time for a single electron density optimization for 128 hydrogen atoms in a given random arrangement at 2 g/cm$^3$. The Kohn-Sham temperature scaling is clearly shown as a prohibitive factor in extending across temperature regimes, while our functional shows no such issue.}
  \label{fig:Tscale}
\end{figure}

First we consider the computational cost. The time required for the optimization of the electron density for a given random arrangement of 128 hydrogen atoms at density 2 g/cm$^3$ at various temperatures is shown in Fig. \ref{fig:Tscale}. This corresponds to the time for a single molecular dynamics time step. For the Kohn-Sham case the temperature scaling is clearly shown as a bottleneck to higher temperature simulations as the time goes from under 10 seconds at 1 eV to over 1200 seconds at 24 eV and over 4300 seconds at 32 eV. The required number of orbitals goes from 100, to 1600, to 2400 respectively to achieve a threshold occupation of $10^{-6}$. The calculation was not pushed above 32 eV on the 8-core 2.93 Ghz Intel Xeon benchmarking machine. In contrast for the orbital-free methods there is no scaling with temperature. Our functional took generally 4-6 seconds whereas the simpler Thomas-Fermi calculations took about 2 seconds. This represents the typical increase in cost we have seen with our functional over the Thomas-Fermi functional. It is of note that though the nonlocal terms of Eqs. (\ref{eq:betavw}) and (\ref{eq:abnl}) appear computationally expensive, they may be evaluated efficiently in reciprocal space through use of fast Fourier transforms (see Supplemental Material \cite{SuppMatt}). 

\begin{figure}[]
  \includegraphics[angle=-90,width=0.9\columnwidth]{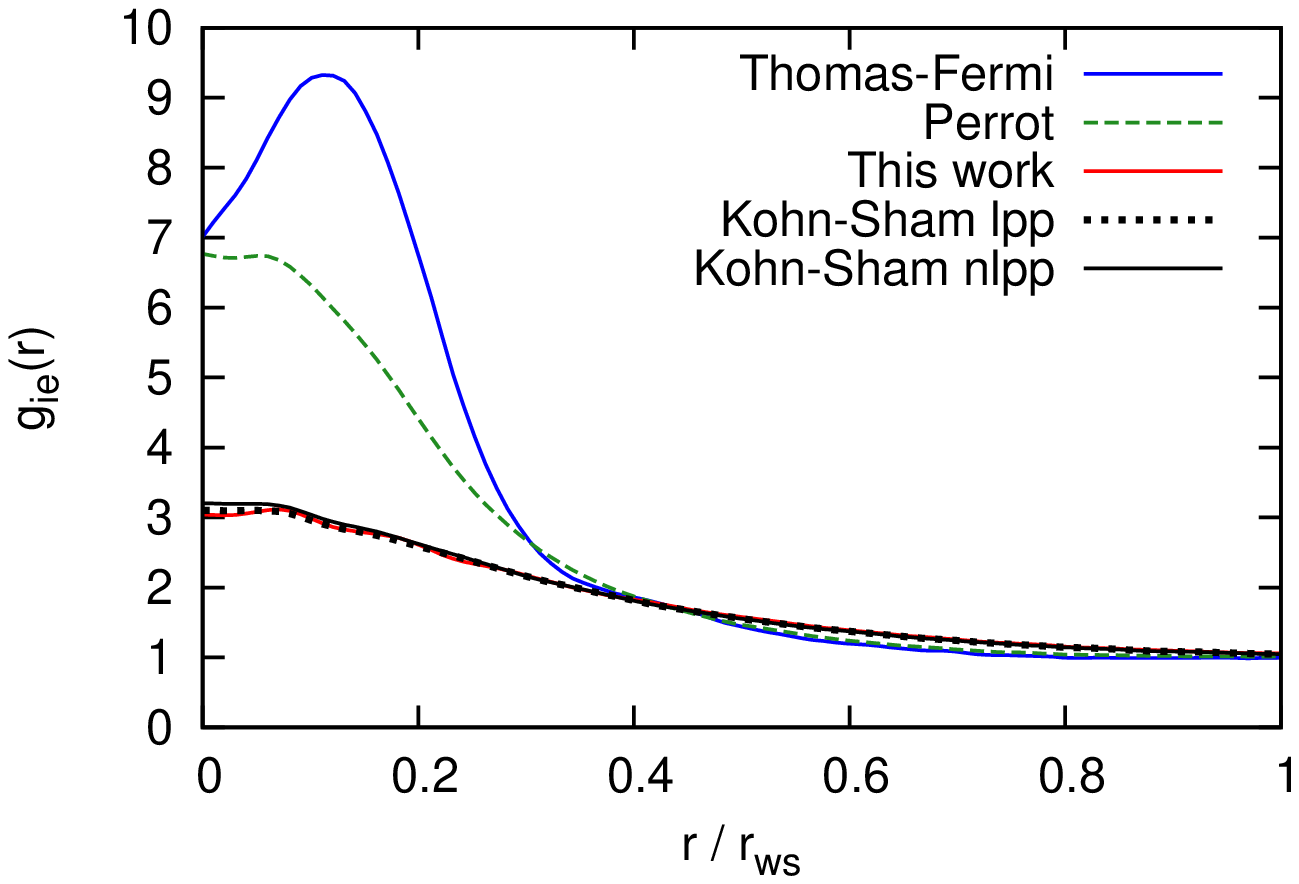}
  \includegraphics[angle=-90,width=0.9\columnwidth]{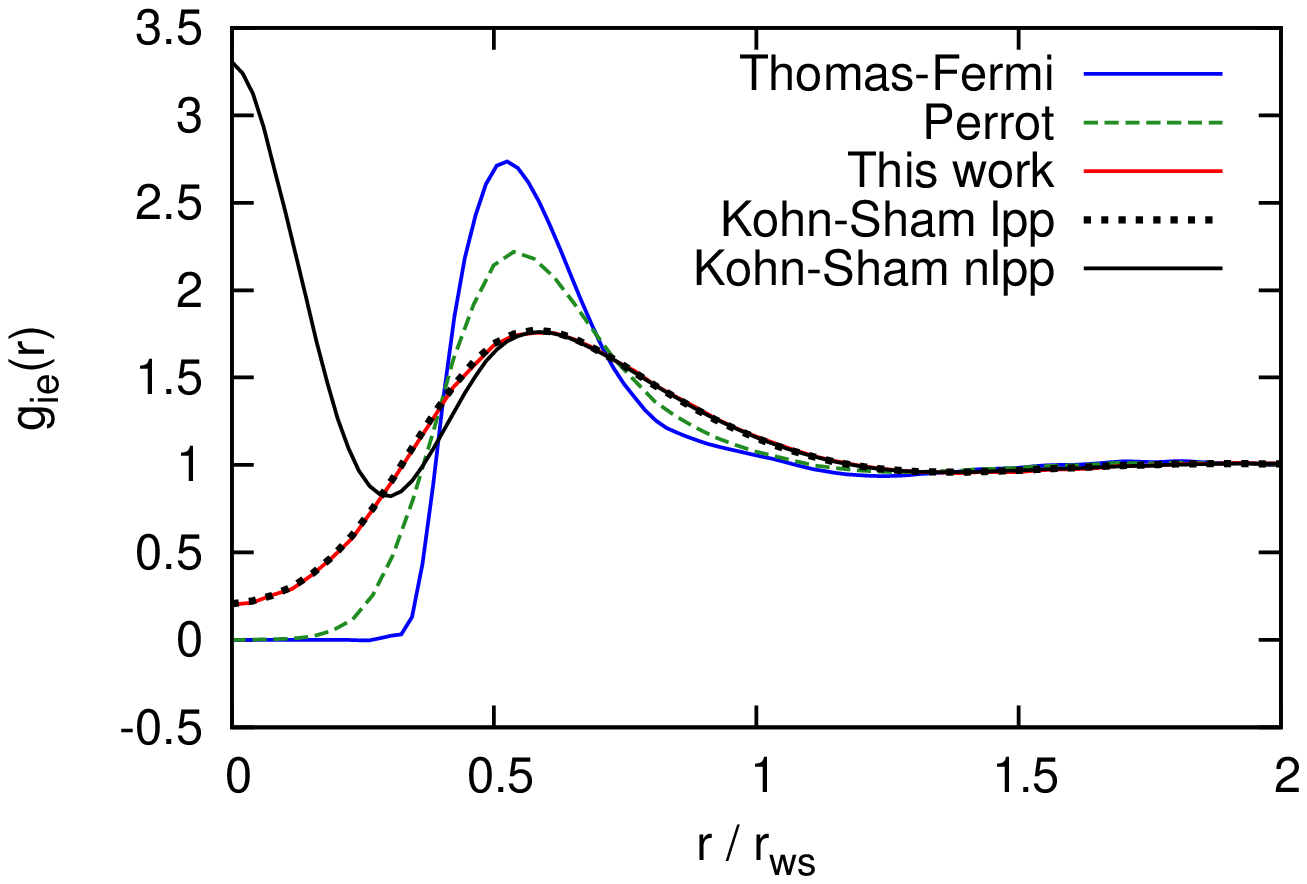}
  \caption{Top: The ion-electron pair correlation function $g_{ie}(r)$ is plotted for a random distribution of hydrogen atoms at density 2 g/cc and temperature of 5 eV. Bottom: is the same but for aluminum at 2.8 g/cc and temperature 100 K = 0.008617 eV. $r_{ws}$ is the ion Wigner-Seitz radius. Both systems show excellent agreement for the electron densities between our functional and Kohn-Sham where the same pseudopotential is used.}
  \label{fig:gie}
\end{figure}

Next we consider an important microscopic feature, the electron density, which by the primary tenant of DFT determines the system completely. Other integrated quantities such as the total energy or pressure, which are often alone considered in determining the accuracy of a functional, are important results. However, if one achieves good results in those integrated quantities and not in the density itself, the integrated results are good due to some cancellation of errors. So we begin with the electron density examined through the ion-electron pair distribution function, $g_{ie}$. Recall that $n(r)=n_0 g_{ie}(r)$ is the average electron density around an ion. In Fig. \ref{fig:gie} the results of three orbital-free functionals are plotted. These include the Thomas-Fermi approximation, as well as the Perrot functional, and our new functional given in this work. As explained before the only difference between these orbital-free calculations and the Kohn-Sham local pseudopotential (lpp) calculation is in $F_s$. The most remarkable feature is that the Kohn-Sham (lpp) and our functional produce nearly identical $g_{ie}$ or electron densities. On the contrary the simpler functionals produce quite different densities. 
We also solve the Kohn-Sham system with the more standard approach of a nonlocal pseudopotential (nlpp). Comparing the Kohn-Sham (nlpp) results we see good agreement for the hydrogen case over the whole range and good agreement for aluminum outside the pseudopotential cutoff radius, around $r/r_{ws} =0.6$. 

\begin{figure}[]
  \includegraphics[width=0.9\columnwidth]{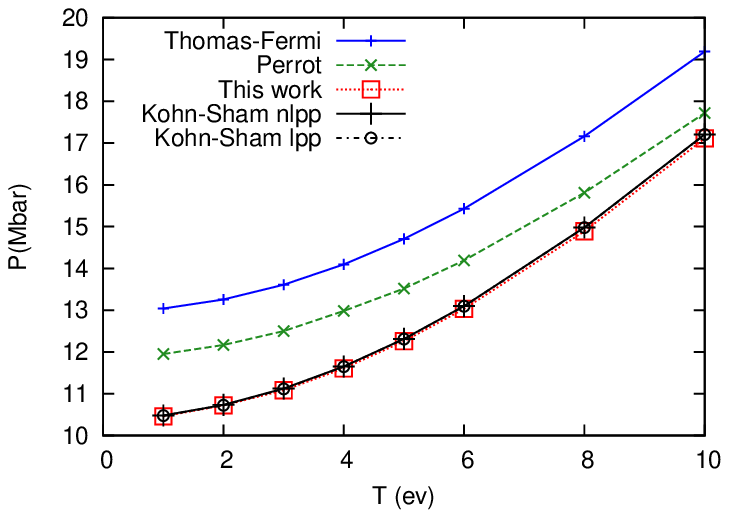}
  \includegraphics[angle=-90,width=0.9\columnwidth]{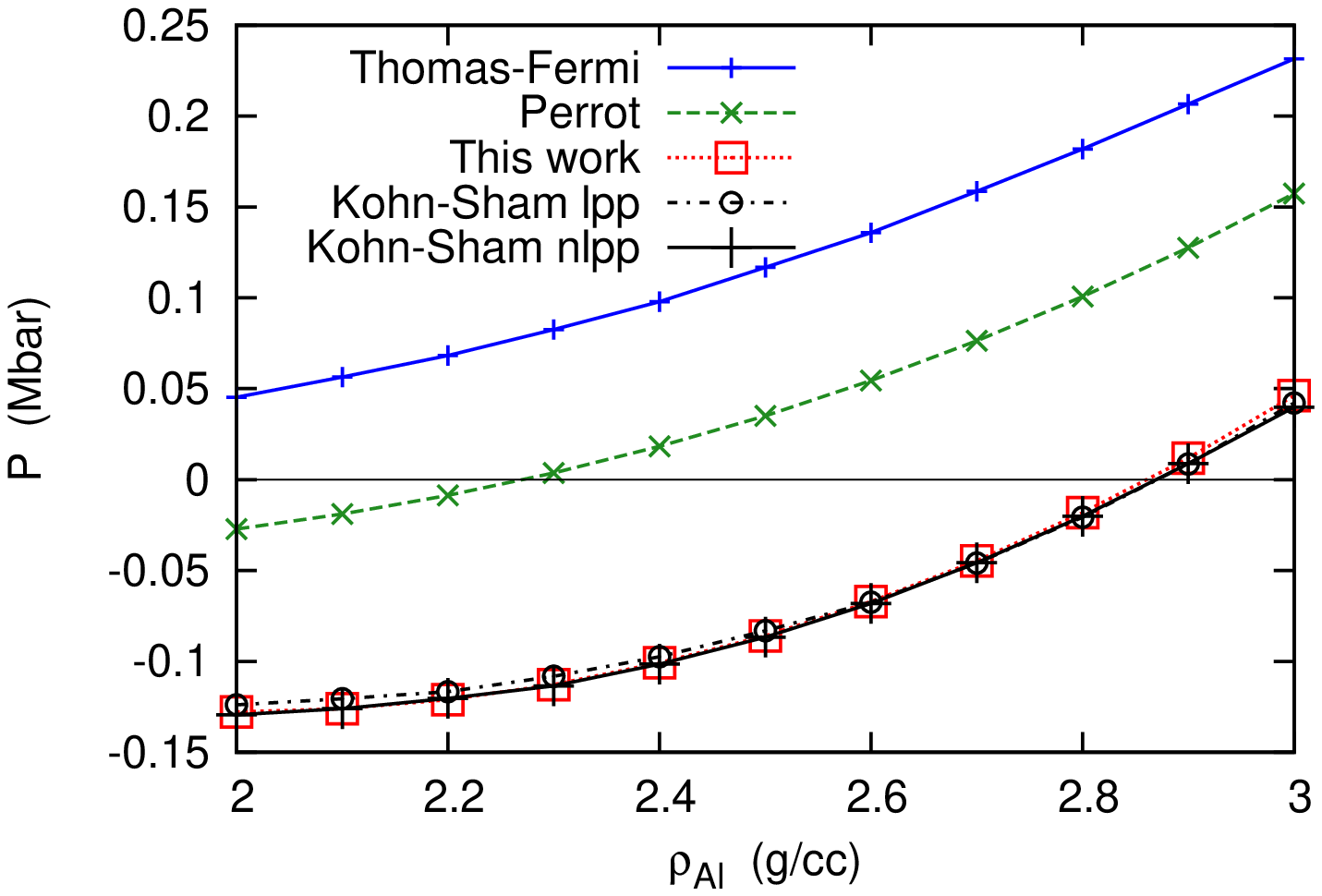}
  \caption{Pressure results for simple cubic hydrogen at 2 g/cc (top) and for fcc aluminum at 100 K = 0.008617 eV (bottom). Both results show our functional closely reproducing Kohn-Sham results.}
  \label{fig:Hsc}
\end{figure}

Next we consider two cases of fixed ions. First we consider hydrogen as a simple-cubic lattice at 2 g/cc and temperatures from 1 to 1000 eV. In the top panel of Fig. \ref{fig:Hsc} the pressure is plotted up to 10 eV for the functionals and pseudopotentials as previously described. The maximum difference between the Kohn-Sham (nlpp) results and our functional is less than 0.5\%, whereas the maximum difference for Thomas-Fermi and Perrot functionals are 24\% and 14\% respectively. Above 40 eV the differences between the functionals is negligible.
In the lower panel we consider face-center-cubic aluminum at 100 K = 0.008617 eV near equilibrium density. Here again there is excellent agreement for the present functional and Kohn-Sham methods. The simple Thomas-Fermi functional does not exhibit any binding, as indicated by the pressure becoming negative, and while the Perrot correction does it is significantly different from the results of Kohn-Sham and our functional.

\begin{figure}[]
  \includegraphics[width=0.9\columnwidth]{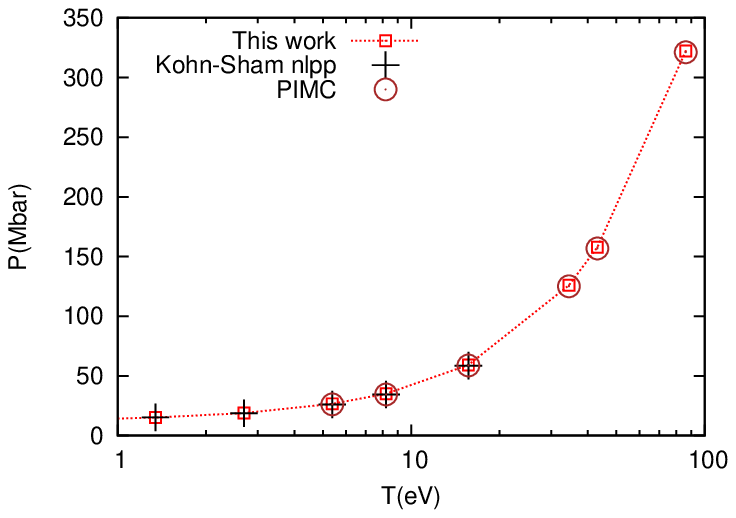}
  \includegraphics[width=0.9\columnwidth]{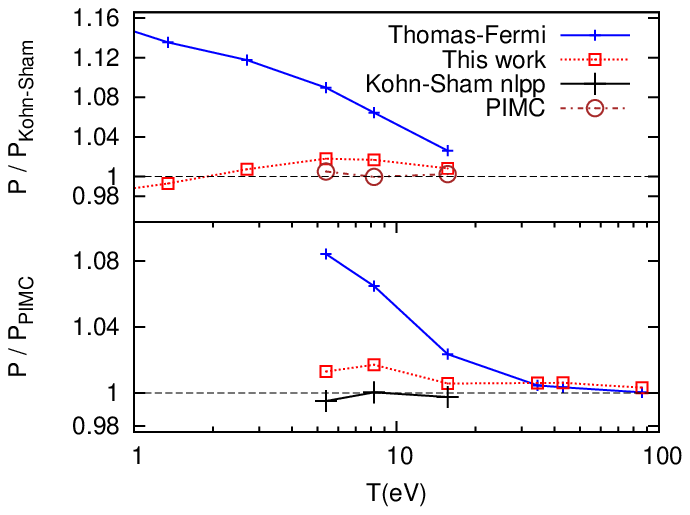}
  \caption{Pressure results for deuterium at 4.04819 g/cc. Our functional is in good agreement with Kohn-Sham and PIMC and spans the entire temperature range. Bottom panel shows the relative pressure with Kohn-Sham and PIMC in their respective ranges.}
  \label{fig:Hmd}
\end{figure}

Now we consider molecular dynamics simulations for warm dense deuterium and aluminum. Note deuterium is examined to connect with the PIMC data, and involves the same pseudopotentials as for hydrogen. Equation of state results are plotted for deuterium at 4.04819 g/cc and temperatures from 1 to 100 eV in Fig. \ref{fig:Hmd}. In addition Kohn-Sham results are plotted up to 15.7 eV and path integral Monte Carlo \cite{Huetal} results down to 5.4 eV. While the Kohn-Sham method becomes computationally prohibitive with increasing temperature the PIMC does so with decreasing temperature. The present orbital-free calculations however span the entire temperature and are significantly less expensive than the other methods at any temperature while showing good agreement with both the Kohn-Sham and PIMC in their respective regions of applicability. Specifically our functional results never deviate by more than 2\% from either the Kohn-Sham or PIMC results. Similar results have been obtained at 1.0 and 10.0 g/cc as well (not shown).

\begin{figure}[]
  \includegraphics[angle=-90,width=0.9\columnwidth]{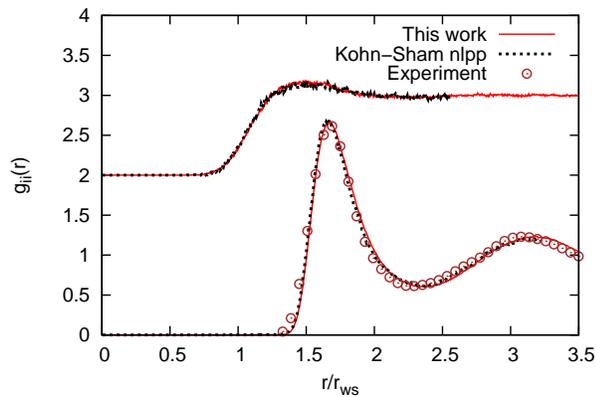}
  \caption{Pair distribution function $g_{ii}(r)$ for Al at experimental density 2.349 g/cc and temperature 1023 K = 0.08815 eV (lower curves) and warm dense conditions 2.7 g/cc and 5 eV (upper curves, shifted by 2). Excellent agreement is shown between Kohn-Sham and our functional.}
  \label{fig:Algii}
\end{figure}

For the case of aluminum we have calculated the ion-ion pair distribution function, $g_{ii}$, for two cases. The first is near melt at the experimental density and temperature of 2.349 g/cc and 1023 K = 0.08815 eV. Second is the warm dense case of 2.7 g/cc and 5 eV. Fig. \ref{fig:Algii} shows $g_{ii}$ for both cases. Our functional and the Kohn-Sham results are in very good agreement in both cases and the experimental data is also in agreement at the lower temperature. 

In summary the present orbital-free functional shows excellent agreement with Kohn-Sham results while being computationally less expensive and having applicability to regions of higher temperature than is accessible by Kohn-Sham methods, as well as very good agreement with PIMC at high temperatures while reaching lower temperatures than accessible by PIMC.  The strong agreement in the $g_{ie}(r)$, as compared with Kohn-Sham method, shows also that the current results are truly reproducing orbital-based results and as such demonstrate a realization of a highly accurate pure density functional theory. In future work we will consider more complex systems with higher atomic number elements as well as mixtures.

\begin{acknowledgments}
  This research has been supported by the DOE Office of Fusion Sciences (FES), and
by the NNSA of the US DOE at Los Alamos National Laboratory under Contract No. DE-AC52-06NA25396. 
\end{acknowledgments}

\clearpage

\appendix

\section{Supplemental Material }
We provide here further details of the three terms of the functional developed in the main paper as well as other quantities for implementation in molecular dynamics simulations. These include the free energy and potential terms as well as the response function, stress tensor and pressure. Forces on the ions only depend on the Coulomb interaction with the electron density and the ions themselves, and are not detailed here.
In general, for below, the response functions are given in the uniform electron gas limit according to Eq. (\ref{eq:linresp}).
Also note that here, $\beta$ shown without any dependency is $1/k_BT$, while with dependence it is the kernel. 

\subsection{Thomas-Fermi}
First for the Thomas-Fermi term
\begin{equation}
  F_{TF}[n] = \int f_{TF}(n(\mathbf{r})) \;d{\mathbf r} \;,
\end{equation}
we provide the free energy density 
\begin{align}
  f_{TF}(n(\mathbf{r}),T) = &\left(\frac{m_e}{\hbar^2}\right)^{3/2} \frac{\sqrt{2}}{\pi^2 \beta^{5/2}} \times 
  \nonumber \\
&\left[ -\frac{2}{3} I_{3/2}(\beta \mu_0) + \beta\mu_0I_{1/2}(\beta \mu_0)\right] 
\end{align}
with the electron density given by
\begin{align}
    n(\mathbf{r}) &= \left(\frac{m_e}{\hbar^2}\right)^{3/2}\frac{\sqrt{2}}{\pi^2 \beta^{3/2}} I_{1/2}(\beta \mu_0)\;,
\end{align}
where the $I_{\nu}$ are Fermi integrals.
The Thomas-Fermi potential is given by the functional derivative, this is just the non-interacting chemical potential
\begin{equation}
  \frac{\delta F_{TF}[n,T]}{\delta n} = \mu_0(n(\mathbf{r})) = \frac{\partial f_{TF}(n,T)}{\partial n}\;.
\end{equation}
The response function is given by
\begin{align}
  \tilde{\chi}_{TF}(T) &= -\left(\frac{m_e}{\hbar^2}\right)^{3/2}
\frac{1}{2\pi^2}\left(\frac{2}{\beta}\right)^{1/2} I_{-1/2}(\beta\mu_0) \label{eq:chitf} \;. \\
\end{align}
The stress tensor is then
\begin{align}
  \sigma_{TF}^{\mu,\nu} = \frac{\delta_{\mu,\nu}}{V} \int f_{TF}(n,T)-n \frac{\partial f_{TF}(n,T)}{\partial n} \;d{\mathbf r}\;.
\end{align}
The pressure is given then by the general expression $P=-\Tr {\sigma}/3$, as
\begin{align}
  P_{TF}=-\frac{1}{V}\int f_{TF}(n,T)-n \frac{\partial f_{TF}(n,T)}{\partial n} \;d{\mathbf r}\;.
\end{align}

\subsection{Nonlocal von Weizs\"acker}
For the von Weizs\"acker term we make use of the root of the density $\phi^2({\mathbf r})=n({\mathbf r})$, noting for use in the chain rule $\delta \phi/\delta n = 1/(2\phi)$. The free energy and potential are 
\begin{align}
 {}_{\beta}F_{vW}[n] = \frac{\hbar^2}{2m_e} \iint& [\nabla \phi(\mathbf{r}))\cdot(\nabla \phi(\mathbf{r}^{\prime})] \times \nonumber \\
& [\delta({\mathbf r}-{\mathbf r}^{\prime})+\beta(|{\mathbf r}-{\mathbf r}^{\prime}|)]\; d{\mathbf r}^{\prime}d{\mathbf r}\;.
\end{align}
\begin{align}
  \frac{\delta {}_{\beta}F_{vW}[n]}{\delta n} = 
-\frac{\hbar^2}{m_e}\frac{1}{2 \phi(\mathbf{r})} \int &[\delta({\mathbf r}-{\mathbf r}^{\prime})+\beta(|{\mathbf r}-{\mathbf r}^{\prime}|)]\times
\nonumber \\
& \nabla^2 \phi(\mathbf{r}^{\prime})\;d{\mathbf r}^{\prime}\;.
\end{align}
In reciprocal space we have for the free energy
\begin{align}
  {}_{\beta}F_{vW}[n] &= \frac{\hbar^2}{2m_e V}\sum_{\mathbf k} (1+\tilde{\beta}(k)) \times \nonumber \\
& \iint [\nabla \phi(\mathbf{r}))\cdot(\nabla \phi(\mathbf{r}^{\prime})] 
e^{-i{\mathbf k}\cdot({\mathbf r}-{\mathbf r}^{\prime})}\; d{\mathbf r}^{\prime}d{\mathbf r}\;. \nonumber \\
&=\frac{\hbar^2}{2m_e V}\sum_{\mathbf k} (1+\tilde{\beta}(k)) (i {\mathbf k}) \tilde{\phi}(-{\mathbf k}) (-i {\mathbf k}) \tilde{\phi}({\mathbf k}) \nonumber \\
&=\frac{\hbar^2}{2m_e V}\sum_{\mathbf k} k^2 (1+\tilde{\beta}(k))\tilde{\phi}(-{\mathbf k}) \tilde{\phi}({\mathbf k})
\label{eq:fvwk}
\end{align}
and the potential
\begin{align}
  \frac{\delta {}_{\beta}F_{vW}[n]}{\delta n} = -\frac{\hbar^2}{m_e}\frac{1}{2 \phi(\mathbf{r})} 
  \hat{\mathrm{F}}^{-1} \left[ k^2 \tilde{\phi}({\mathbf k}) (1+\tilde{\beta}(k)) \right] \;.
\end{align}
The response function is then
\begin{align}
  {}_{\beta}\tilde{\chi}_{vW}(\mathbf{q},T) &= -\frac{m_e}{\hbar^2}\frac{4 k_F^3 }{ 3\pi^2k^2} 
\left(\frac{1}{1+\tilde{\beta}(k)}\right) \;. \\
\end{align}
For the stress tensor it is important to note that $\beta(k)$ is actually a function of $q=k/k_F$ only, so that
\begin{align}
  {}_{\beta}\sigma_{vW}^{\mu,\nu} &=-\frac{\hbar^2}{m_e}\frac{1}{2 V^2}
  \sum_{\mathbf k \ne 0} \tilde{\phi}(-{\mathbf k}) \tilde{\phi}({\mathbf k}) \times 
\nonumber \\
& \left\{  k_{\mu}k_{\nu} 2 (1+\tilde{\beta}(q)) + k^2 \frac{\partial (1+\tilde{\beta}(q))}{\partial q} \frac{\partial q}{\partial \epsilon_{\mu,\nu}}\bigg|_{\epsilon=0}\; \right\}.
\end{align}
This then yields for the pressure
\begin{align}
  {}_{\beta}P_{vW} &=\frac{\hbar^2}{m_e}\frac{1}{2 V^2}
\sum_{\mathbf k \ne 0} \tilde{\phi}(-{\mathbf k}) \tilde{\phi}({\mathbf k}) \frac{2}{3} k^2 (1+\tilde{\beta}(q)) \nonumber \\
&= \frac{2}{3V} {}_{\beta}F_{vW}
\end{align}
The simplification comes from the use of the dimensionless $q$ since
\begin{align}
  \frac{\partial q}{\partial \epsilon_{\mu,\nu}}\bigg|_{\epsilon=0}&=  \left( \frac{\partial q}{\partial k}\frac{\partial k}{\partial \epsilon_{\mu,\nu}}\bigg|_{\epsilon=0} + \frac{\partial q}{\partial k_F} \frac{\partial k_F}{\partial \epsilon_{\mu,\nu}}\bigg|_{\epsilon=0}\right) \nonumber \\
&=\left( -\frac{k_{\mu}k_{\nu}}{k k_F} + \frac{k}{3 k_F}\delta_{\mu,\nu} \right) \;,
\label{eq:dqde}
\end{align}
where the terms in the parentheses will give zero contribution to the pressure in the trace.

\subsection{Nonlocal density term}
Finally for $F_{a,b}$ 
\begin{equation}
  F_{a,b}[n] = \iint n^{a}(\mathbf{r}) w(\mathbf{r}-\mathbf{r^{\prime}},T) n^{b}(\mathbf{r}^{\prime}) \;d\mathbf{r}\;d\mathbf{r}^{\prime} \;.
\end{equation}
the functional derivative may be immediately taken to find 
\begin{align}
  \frac{\delta F_{a,b}[n]}{\delta n} =&  an^{a-1}(\mathbf{r}) \int w(\mathbf{r}-\mathbf{r^{\prime}},T) n^{b}(\mathbf{r}^{\prime}) \;d\mathbf{r}^{\prime} \nonumber \\
  &+ bn^{b-1}(\mathbf{r}) \int w(\mathbf{r}-\mathbf{r^{\prime}},T) n^{a}(\mathbf{r}^{\prime}) \;d\mathbf{r^{\prime}} \;.
\end{align}
It is convenient for implementation to have the stress tensor in reciprocal space, so first we rewrite the free energy component as
\begin{align}
  F_{a,b}[n]= \frac{1}{V} \sum_{\mathbf k} \tilde{w}(k) \tilde{n}_a(-\mathbf{k}) \tilde{n}_b({\mathbf k})
\end{align}
or
\begin{equation}
  F_{a,b}[n] = \int n^a(\mathbf{r},T) \hat{\mathrm{F}}^{-1} \left[
    \tilde{w}(k) \tilde{n}_b(\mathbf{k}) \right] \; d\mathbf{r} \;.
\end{equation}
where $\tilde{w}(k)$, $\tilde{n}_a(\mathbf{k})$, $\tilde{n}_b({\mathbf k})$ are the respective Fourier transforms of $w(\mathbf{r}-\mathbf{r^\prime})$, $n^a(\mathbf{r})$, $n^b(\mathbf{r})$. The potential is similarly
\begin{align}
  \frac{\delta F_{a,b}[n]}{\delta n} =&
   an^{a-1}(\mathbf{r}) \hat{\mathrm F}^{-1} \left[ \tilde{w}(k) \tilde{n}_b({\mathbf k})\right] \nonumber \\
   & + bn^{b-1}(\mathbf{r}) \hat{\mathrm F}^{-1} \left[ \tilde{w}(k) \tilde{n}_a({\mathbf k})\right] 
\end{align}
Here for the stress tensor we note $\tilde{w}$ is a separable function of $q$ and $n_0$ with the $n_0$ dependence just coming from the coefficient as
\begin{gather}
  \tilde{w}(k) = u(q) W(n_0) \nonumber \\
  W(n_0)=\frac{1}{2 a b n_0^{(a+b+1/3-2)}} \frac{\hbar^2 \pi^2}{(3\pi^2)m_e}\;.
  \label{eq:uq}
\end{gather}
The stress tensor is then
\begin{align}
   \sigma_{a,b}^{\mu,\nu} = &\frac{1-(a+b)}{V}F_{a,b}\delta_{\mu,\nu} \nonumber \\
&+ \frac{1}{V^2} \sum_{\mathbf{k}\ne 0}  \tilde{n}_a(\mathbf{k}) \tilde{n}_b(-\mathbf{k}) 
\times \nonumber \\
&\bigg[(a+b+\frac{1}{3}-2)\frac{u(q)}{W(n_0)} \delta_{\mu,\nu} \nonumber \\
&+\frac{1}{W(n_0)}\frac{\partial u(q)}{\partial q}\frac{\partial q}{\partial \epsilon_{\mu,\nu}}\bigg|_{\epsilon=0}
\bigg]\;.
\end{align}
The pressure follows as
\begin{align}
  P_{a,b} =& \frac{2}{3V} F_{a,b} \;.
\end{align}
And the response is
\begin{align}
\tilde{\chi}_{a,b}(\mathbf{k})^{-1} 
&= -\left( 2 a b n_0^{(a + b - 2)} \tilde{w}(k) \right)
\label{eq:chilr}
\end{align}
with $\tilde{w}(k)$ found through Eq. (\ref{eq:w}). This result requires $\tilde{w}(k=0)=0$, which we enforced in Eq. (\ref{eq:w}) by taking $f(k)$ as finite.

\end{document}